# Data-Driven Model For Heat Load Prediction In Buildings Connected To District Heating Networks

Alaeddine Hajri[1], Roberto Garay-Marinez[2, *], Ana Maria Macarulla[2], Mohamed Amin Ben Sassi[1]

[1] Department of Computer Systems Engineering, Mediterranean Institute of Technology, Tunis, Tunisia

[2] Institute of Technology, Faculty of Engineering, University of Deusto, Avda. Universidades, 24, Bilbao 48007, Spain

* Roberto.garay@deusto.es, Tel. +34 94 413 90 00

## Abstract

In this study we investigate the heat load patterns in one building using multi-step forecasting model. We combine the Autoregressive models that use multiple eXogenous variables (ARX) with Seasonally adaptable Time of Week and Climate dependent models (S-TOW-C) (to correct model inaccuracies), to obtain a robust and accurate regression model that we called S-TOW-C-ARX used in time series forecasting.

Based on the experiment results, it has been shown that the proposed model is suitable for short term heat load forecasting. The best forecasting performance is achieved in winter term where the prediction values are from 4 to 20 % away from the targets, which are commonly seen as very good values.

# Nomenclature

| | |
|---|---|
| DH | District Heating |
| SARIMA | Seasonal Autoregressive Integrated Moving Average |
| OLS | Ordinary Least Square |
| SVR | Support Vector Regression |
| MPR | Multilayer Perceptron |
| MAPE | Mean Absolute Percentage Error |
| DHW | Domestic Heat Water |
| IQR | Interquartile Range |
| ARX | Autoregressive with Exogenous variables |
| S-TOW-C-ARX | Seasonally Adaptable, Time of Week and Climate Dependent Auto Regressive Model |
| $a_{na}$ | Autoregressive Coefficient |
| $b_{nb}\ c_{nc}$ | Exogenous Coefficient |
| $e_{ni}$ | Calendar Data Coefficient |
| $L_i$ | Heat Load at Time t |
| $L_{i-n_a}$ | Lagged Heat Load |
| $T_{i-n_b}$ | Lagged Temperature Variable |
| $I_{i-n_c}$ | Lagged Solar Irradiation |
| $bool\_CD$ | Boolean Calendar Data |
| $w_i$ | White Noise |
| c | Constant |

# 1 Introduction

Thermal energy uses in buildings form a large part of global energy demand. 50% of Europe's total gross final energy consumption are destined for heating and cooling. According to 2018 Eurostat data [1], where the lion's part (79%) corresponds to heating and domestic hot water production at household level.

In large buildings, heating systems are typically centralized. Also, in many cities, District Heating (DH) systems are used to deliver heat to large sections of the cities. Unlike boilers and heat pumps, DH can use a wide variety of energy sources, including renewable energy uses as well as fossil fuels. [2]

Finding an overall balanced approach to deal with the robust design of heating systems, along with their efficiency, is one of the central dilemmas for building engineers. In this context, these systems are sized to meet the peak demand rather than focusing on their optimal efficiency.

The control for supply temperature of buildings as well as DH networks can improve the efficiency of these systems [3]. Usually, these controls are static and rule-based, they are commonly based on simplified thermostatic rules. These rules could be optimized through a suitable adaptation based on the evolution of the heat demand in the coming hours. Even basic temperature optimization has led to reductions of 4.8% of distribution loss in real life cases.[4]

Our study will focus on how to characterize and forecast the heating demand to ensure the effective operation of large buildings systems. For doing so, we use data driven modelling approach, where the consideration of physical background of the building is not required. So-called black box modelling is a fast and efficient way of creating load models [5]. By using data-driven approaches, it becomes possible to make improvements and adjustments over time by allowing for adaptation and updating of the model based on new data.

In this regard, several studies have been carried out using these approaches. The study of Eguizabal et al., (2021) [6] proposed a data driven model based on present and recent history of weather data. This type of model, so called AutoRegressive with eXogenous variables (ARX), was optimized at 4-hour historical values. The model was tested against synthetic data of a building energy simulation, with forecasting error lower than 4% in winter term.

Lumbreras et al., (2021) [7] developed a Time-of-Week (ToW) Segmented Multiple Linear Regression (MLR) model considering several climatic variables. The model delivered overall fine performance at an aggregated level, but somehow failed at replicating intra daily patterns. Potentially due to its lack of history.

Grosswindhager et al., (2011) [8] proposes a seasonal autoregressive integrated moving average (SARIMA) process for modelling the system heat load. Although suitable to replicate yearly patterns, this model lacks the capacity to adapt to variations in present weather conditions, as it lacks exogenous climatic variables that may have an impact on the heat load.

Several [7], [9], [10] have showed the relevance of calendar and social data in the modelling of heat loads in buildings. Three different machine learning models, namely ordinary least square (OLS), support vector regression (SVR) and multilayer perceptron model (MPR) are proposed by Dahl et al., (2018) [11] to investigate the inclusion of calendar data and holiday data for forecasting the aggregated heat load of the DH system of Aarhus, Denmark. The SVR show the best performance with an MAPE of 6.4%, nevertheless, the inclusion of local holiday showed a minor overall improvement. This indicates the limited interpretability of the proposed model.

For the purpose of this study, we introduce an AutoRegressive model with Multiple eXogenous variables. We incorporate also Time of the Week (ToW) variables as eXogenous variables. By doing so, we believe that it allows to incorporate ToW sensitivity over an otherwise stable ARX model, and we avoid the need for segmentation. This is a mixture of the approaches in the above-mentioned [6], [7] leading to a potentially interesting application. Based on a common approach, we develop various adaptations for specific periods of the year. Ultimately, the model is capable of modelling the unknown with a minimum number of parameters, achieving substantial simplification of linear prediction and providing an effective linear model of stationary time series.

## 2 Methodology

In this section we describe our data-driven approach. This approach is outlined in figure 1. It involves four major steps. (1) Data sources (2) Outlier detection and data imputation. (3) Model definition. (4) Model calibration.

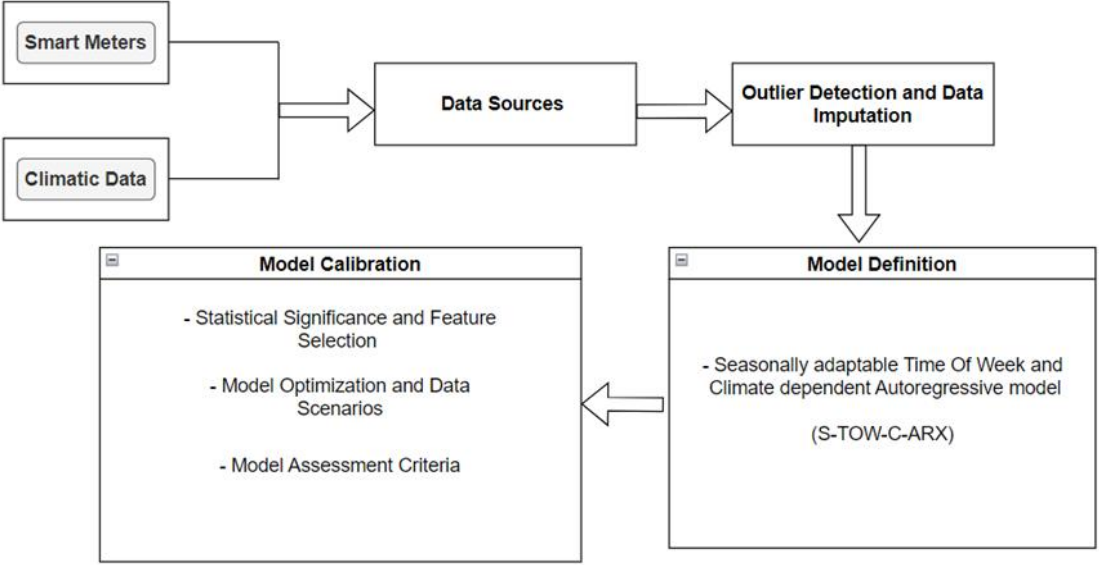

**Figure 1. General Methodology of the study**

## 2.1 Data Sources

Two sources of data are used: heat load data and climate data. The sub-network of the district heating covers 42 substations (buildings), located in Tartu (Estonia). The datasets were collected from each substation. We attached, in each building, an energy smart meter (Kamstrup) [12] that measures constantly different variables. The device sends the real time measured data in hourly basis to GREN, the operator responsible for the district heating [13]. Data taken from the weather station is located in the university of Tartu and have 15-minute frequency basis [14], [15], [16], [17].

For the selected building, we have performed the load assessment process in two different periods: winter (January- February-December) and shoulder (March-April-May+September-October-November) to perceive if there are any hidden patterns.

The increase of temperature in figure 2 (a) and (b) below indicates the strong correlation between weather variable and heating demand. Part of the heating demand is dedicated to domestic hot water consumption (DHW) which is illustrated by the roughly horizontal profile against the weather data as shown figure 2 (b). Indeed, the hot water consumption shows little to no dependence on weather variable and seasonal variation.

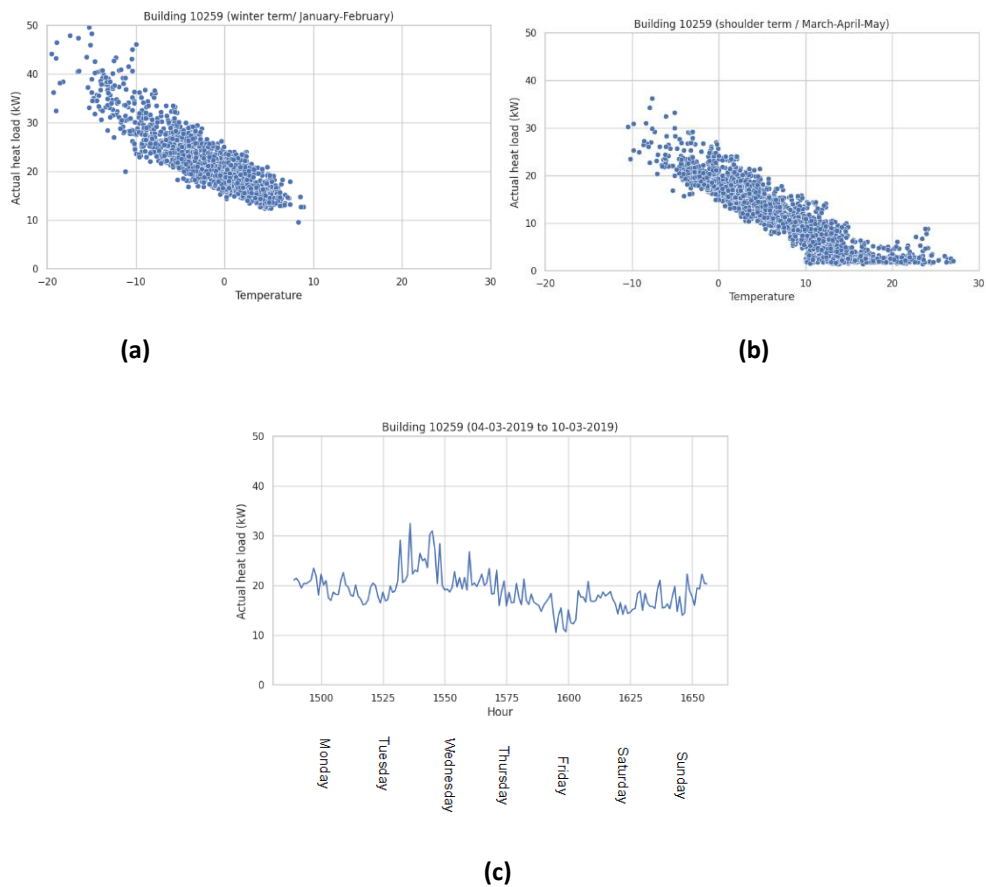

Figure 2. Heat load vs outdoor temperature in Building 10259: Winter term (a) and Shoulder term (b) with Weekly Consumption (04-03-2019 to 10-03-2019) (c)

Multiple seasonality is observed, where the weekly patterns reflect the variation in heat load on weekdays versus weekends. The weekly seasonality is exemplified in fig 2 (c) above, representing the heat load consumption for one week in shoulder term. The decrease of heat load at the end of the week could be explained by the holidays or other exceptional events.

## 2.2 Outlier Detection and Data Imputation

Outliers have been detected and removed based on statistical significance tests. Considering the sensitivity of data to climate, the process was conducted individually for temperature-homogeneous areas at intervals of 2.5°C. Anomalies were identified using the IQR rule for each subset of temperature. All observation lying above the third quartile + 1.5*IQR and lying below the first quartile - 1.5*IQR, where IQR = third quartile - first quartile, are considered outliers. Based on the defined outlier zone, a total of 8.42% of the data points were identified.

## 2.3 Model Definition

Our proposed model uses Seasonally adaptable, time of week and climate dependent Auto Regressive Model" (S-TOW-C-ARX). It is expected to forecast short- term heat load with hourly load data. The model adapts through discrete ($e_{ne}$) deviations of the load for specific TOW values. The linear regression function is performed in order to identify the coefficients, where the heat load output value is considered in previous timesteps. The model components are presented in the following equation:

**Equation 1. (S-TOW-C-ARX) Model Structure**

$$L_i = c + \sum(a_{na} \times L_{i-na}) + \sum(b_{nb} \times T_{i-nb}) + \sum(c_{nc} \times I_{i-nc}) + \sum(e_{ne} \times bool\_CD) + w_i$$

Where $a_{na}$ are the autoregressive parameter, $b_{nb}$, $c_{nc}$ are the exogenous parameters corresponding to Temperature and Solar Irradiation, and $e_{ni}$ are the parameter related to calendar data. The $L_i$ refers to the heat load at time t (hours), $L_{i-na}$, $T_{i-nb}$ and $I_{i-nc}$ represent the present and lagged heat load, temperature and irradiation flux variables respectively. The calendar data are symbolized as $bool\_CD$ which represent each hour of the week separately. Finally, $w_i$ is identified as white noise parameter with a constant term c as well.

## 2.4 Model Calibration

The dataset was split into training and testing sets. The splitting process was performed in a way that the learning data and testing data coincide with all different heat consumption patterns with the respect to the season of the year. The training data have been defined containing odd days whereas the testing set have been defined containing even days.

The data have been limited only to winter term (December to February) and shoulder season (March - May + Sep - Nov). Data from the summer was not considered relevant due to its low load values. Furthermore, the workdays and weekends were separated in order to remove the intraweek seasonality.

### 2.4.1 Statistical Significance and Feature Selection

Models were fitted to the ordinary least square regression (OLS). Thereafter feature selection was conducted using variance threshold. The selection procedure was based on p-value threshold in which the entry criterion was set to a value lower than 0.05.

This procedure was complemented with visual checks of the accuracy of the models after the feature selection to ensure that our assumptions and feature selections were reasonable

### 2.4.2 Model Optimization and Data Scenarios

To evaluate the effect of including various types of input variables to predict the heat demand, three different types of data (called " data scenario ") are demonstrated. In other words, the models are built with different available explanatory variables. Table 1 below displays the different Data Scenarios where several variables are added to the models based on an iterative process using forward selection method. In each model, optimal input parameters are selected in order to achieve the best performance of prediction accuracy of heat load.

**Table 1. Models with different data scenarios.**

|  | 12 historical values of Power | Present Temperature | 24 historical values of Temperature | Present Irradiation | 24 historical values of Irradiation | Calendar Data |
|---|---|---|---|---|---|---|
| lagged heat load data | ✓ |  |  |  |  |  |
| lagged weather data (Temperature) | ✓ | ✓ | ✓ |  |  |  |
| lagged weather data (Irradiation flux) | ✓ | ✓ | ✓ | ✓ | ✓ |  |
| calendar data | ✓ | ✓ | ✓ | ✓ | ✓ | ✓ |

### 2.4.3 Model Assessment Criteria

Modelling accuracy is tested with training and testing datasets.

First, using the training data, the accuracy of the models is measured in order to characterize the heat load. In this context, we have checked the goodness of fit for each model with the calculation of R-squared value and F-statistic. Furthermore, a graphical visual inspection of residual analysis is conducted.

Second, using the testing data, the accuracy of the models is measured in order to predict the heat load. The selection of optimal input parameters is evaluated using different evaluation metrics such as the Akaike information criterion (AIC), Bayesian information criterion (BIC)and Adjusted R-squared.

The models with their data scenarios are scored according to hourly mean absolute error (MAE), root mean square error (RMSE) and mean absolute percentage error (MAPE).

## 3 Results and Discussion

## 3.1 Model Evaluation to Characterize the Heat Load

### OLS Goodness of Fit

Table 2. Summary statistics with significant parameters

| Evaluation Metrics | Winter Term | | Shoulder Term | |
|---|---|---|---|---|
| | Workdays | Weekends | Workdays | Weekends |
| R-Squared | 0.845 | 0.763 | 0.883 | 0.866 |
| F-statistic | 342.7 | 76.03 | 707 | 359.7 |
| Prob (F-statistic) | 8.52e-296 | 5.68e-80 | 0.00 | 1.17e-258 |

As shown in table 2 above, the shoulder season models yield an excellent fit to the observed data. For workdays models 88% of the variability observed in the heat load variable is explained by the regression model. The minimum value of R-squared is shown in the winter term, essentially for weekends model where 76 % of the variability observed in the heat load variable is explained by the regression model.

The models of the shoulder term displayed the higher F-statistic values. For the prob (F statistic) we could see that all the values were exceedingly small numbers. For that reason, the null hypothesis of the models was rejected.

## Residual Analysis

Figure 3 (a) (b) below represents the heat load predicted values against the observed values of workdays and weekends models in winter and shoulder terms. In the first subsection (15 to 25 kWh), the models' accuracy does not seem bad. The dispersion of points in the second subsection (30 to 40 kWh) indicates the inadequacy of the models when having high values of heat load in winter terms. Figure 3 (c) and (d)     represents the residuals values against the predicted data. We can see a symmetrically distribution where the values are clustered toward the middle of the plot. However, in shoulder term, we observe a structural pattern particularly in the first subsection. This pattern suggests that the models, in this term, have a room for improvement.

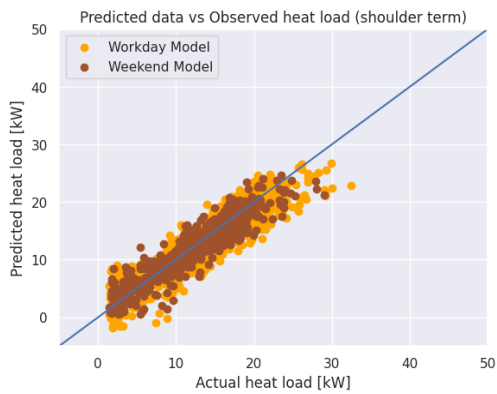
(a)

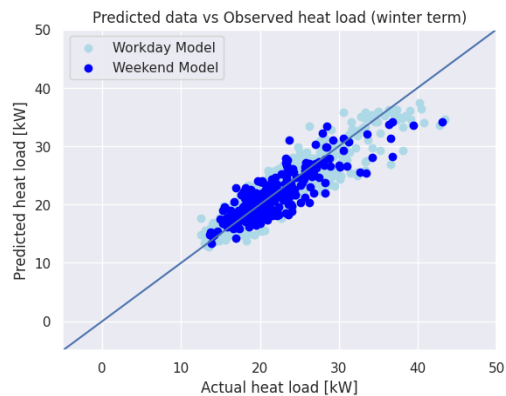
(b)

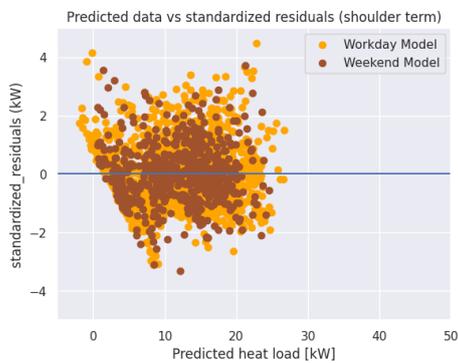
(c)

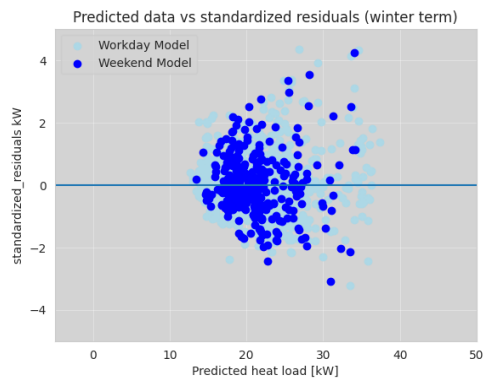
(d)

**Figure 3. Examining the predicted heat load against actual heat load and residuals in winter term (b,d) and shoulder term (a,c)**

## 3.2 Model Evaluation to Predict the Heat Load

### Model Comparison and Selection

From figure 4 below, we observe some variation between the magnitude of the mean absolute error (MAE) and the root mean square errors (RMSE).

Models with only lagged heat load variables stands out by performing significantly worse than models with other data scenarios (lagged weather variables and significant calendar data).

In this perspective, winter models have the highest values of RMSE and MAE compared to shoulder models in all the three data scenarios. (Particularly the weekend model in winter terms with an RMSE value of 3.254 kW and an MAE value of 2.334 kW).

Adding lagged weather variables of temperature and irradiation flux drastically increase the performance of all the four models which is explained by the decreasing value of the mean absolute error (MAE) and the root mean square error (RMSE).

One can see in both figures that adding the significant calendar data slightly improve the performance of the models. The focus in the rest of this work will be on the four models of the 3rd data scenario using significant calendar data.

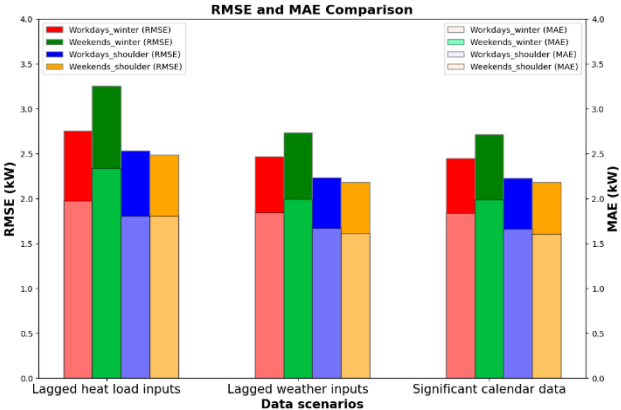

**Figure 4. Root mean square error (RMSE) and mean absolute error (MAE) of the four forecast models (workdays-winter, weekends-winter, workdays-shoulder and weekends-shoulder) for each data scenario**

**Table 3. Equations of the models that include significant calendar data inputs (3rd data scenario)**

| Winter Term | Workday model | $Q = C_0 + C_{Q1} \times Q_1 + C_{Q2} \times Q_2 + C_{Q3} \times Q_3 + C_{T0} \times T_0 + C_{T1} \times T_1 + C_{I0} \times I_0 + C_{I1} \times I_1 + w_i + TUE_{1h} + WED_{3h} + WED_{7h} + MON_{8h}$ |
|---|---|---|
| | Weekend model | $Q = C_0 + C_{Q1} \times Q_1 + C_{T0} \times T_0 + C_{T1} \times T_1 + C_{I0} \times I_0 + C_{I1} \times I_1 + w_i + SAT_{2h} + SAT_{10h} + SUN_{1h} + SUN_{4h} + SUN_{5h}$ |
| Shoulder Term | Workday model | $Q = C_0 + C_{Q1} \times Q_1 + C_{Q2} \times Q_2 + C_{Q3} \times Q_3 + C_{T0} \times T_0 + C_{T1} \times T_1 + C_{T2} \times T_2 + C_{I0} \times I_0 + C_{I1} \times I_1 + C_{I2} \times I_2 + w_i + THU_{2h} + FRI_{7h} + FRI_{10h} + FRI_{11h} + FRI_{21h} + MON_{0h} + MON_{1h} + MON_{8h}$ |
| | Weekend model | $Q = C_0 + C_{Q1} \times Q_1 + C_{Q2} \times Q_2 + C_{Q3} \times Q_3 + C_{T0} \times T_0 + C_{T1} \times T_1 + C_{I0} \times I_0 + C_{I1} \times I_1 + w_i + SAT_{22h} + SUN_{15h} + SUN_{16h} + SUN_{18h} + SUN_{19h} + SUN_{22h}$ |

Table 3 represents the models of the 3rd data scenario which include significant calendar data inputs. The workdays model in winter term is denoted by Q3_T1_I1 (TUE_1h_WED_3,7h_MON_8h): Q3 represents the heat load with three lagged values, T1, I1 represent the lagged values at 1-hour historical data for temperature and irradiation flux respectively. WED_3,7h refers to the significant predictors of hours at 03:00 and 7:00 on Wednesday, MON_8h refers to the significant predictors of hours 08:00 on Monday.

The same notation was used for the other models. The weekends model in winter term is denoted by Q1_T1_I1 (SAT_2,10h_SUN_1,4,5h). The workdays model in shoulder term is denoted by Q3_T2_I2 (THU_2h_FRI_7,10,11,21h_MON_0,1,8h). The weekends model in shoulder term is denoted by Q3_T1_I1 (SAT_22h_SUN_15,16,18,19,22h).

## Seasonal Performance Variation

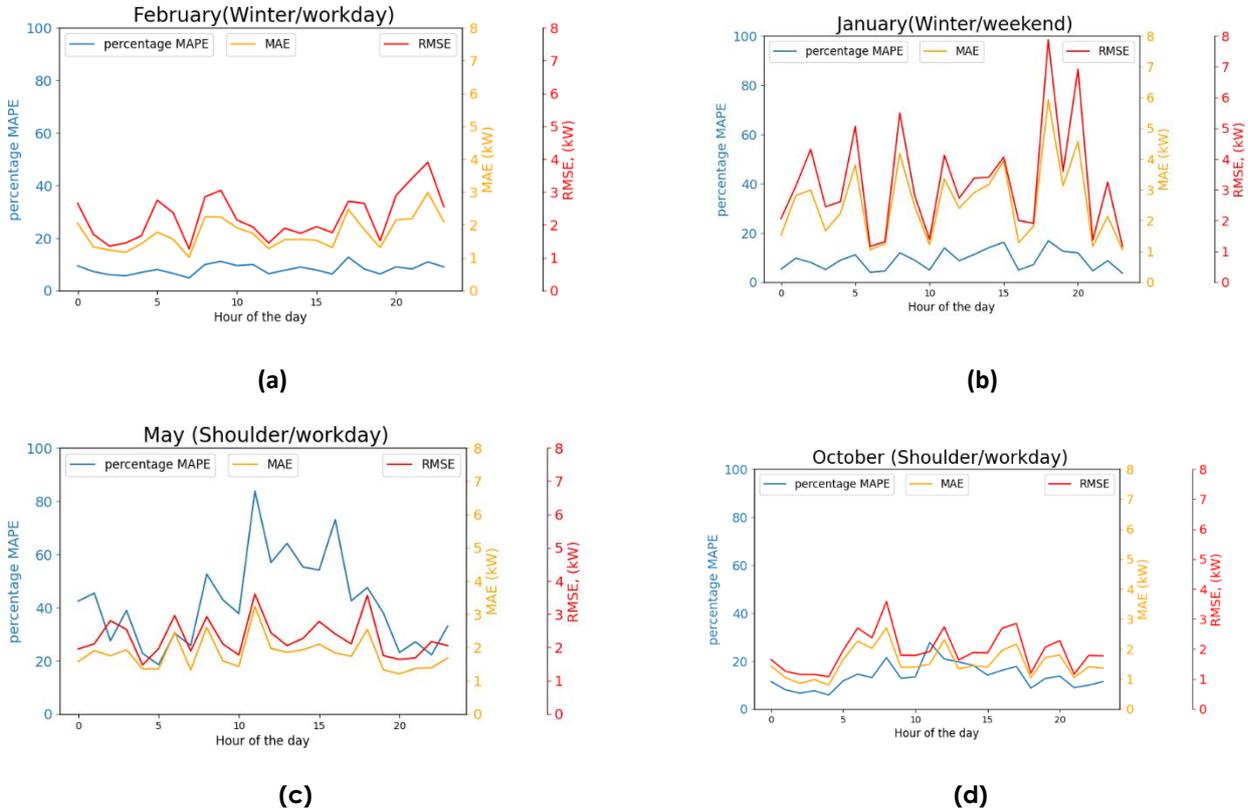

**Figure 5. Examining the performance of the four models for February (a), January (b), May (c), October (d), where the forecast error varies with time of the days.**

As shown in figure 5 above the heat load varies over the year in magnitude and variation. Three different evaluation metrics are shown: on the left axis the mean absolute percentage error (MAPE), on the right axis the mean absolute error (MAE) and root mean square error (RMSE). The hours of the day are represented by the horizontal axis.

We can see that the RMSE and MAE appear to be large in winter term compared to shoulder months. In this context the evaluation metrics are large in the evening hours, this could be explained by the large heat load with large variance during the winter months. The RMSE and MAE can be above 4 kW and 3 kW in some hours especially in January whereas in shoulder term it can be below 2 kW especially in May.

The MAPE behaves in opposite way compared to RMSE and MAE metrics, generally the aforementioned metric is smaller in winter months with a value between 4 to 20 % and larger in shoulder months with values between 20 to 80 %. These large values of MAPE indicate the inaccuracy of forecasting in some hours of shoulder months especially for April, May and September where we observe a peak of MAPE particularly between 10:00 to 15:00.

As a whole we didn't spot any clear pattern of error change during the day. Moreover, the performance of the four models doesn't seem to be bad in early morning particularly between 3:00 to 4:00.

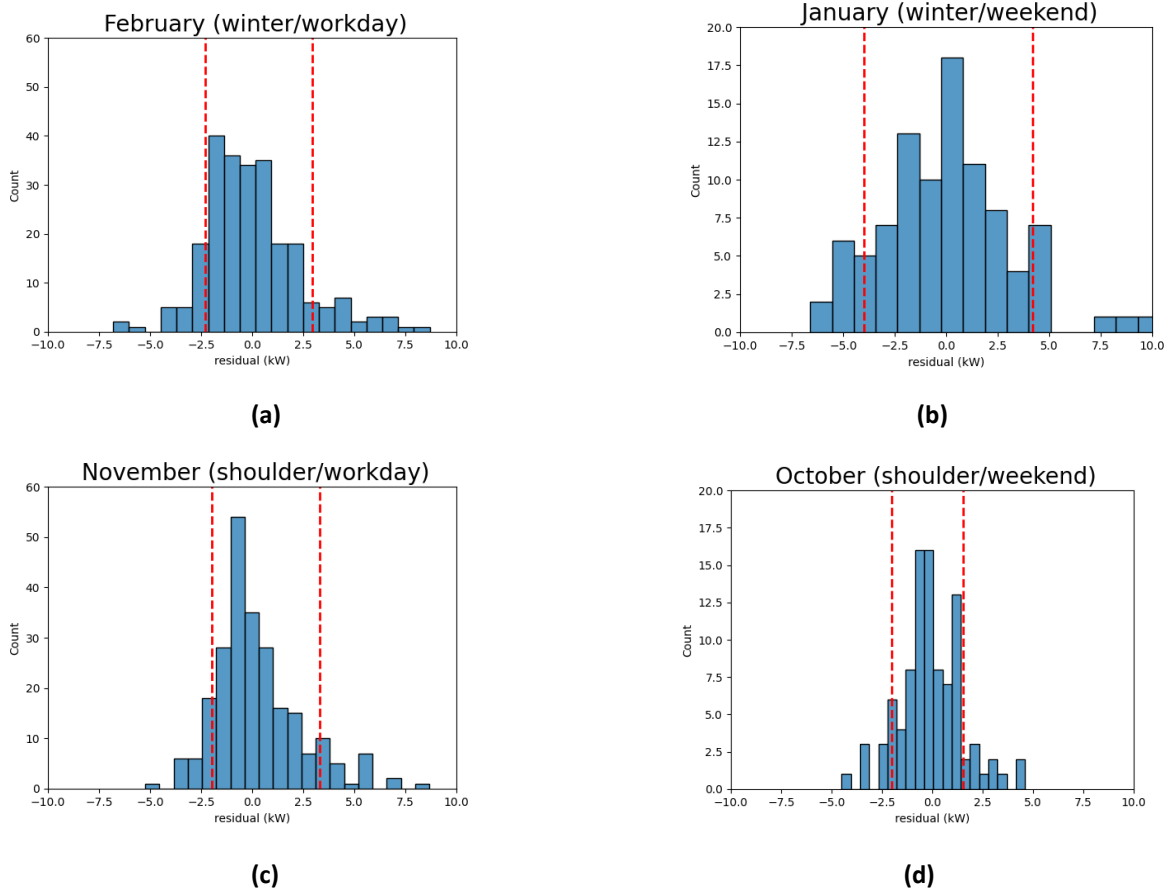

Figure 6. Forecast error histograms of the four models. The forecast error distribution is shown for February (a), January (b), November (c), October (d) in the year along with 10% and 90% quantiles. The forecast is too high when having a positive error, and too low when having negative error

The aggregated error metrics such as RMSE, MAE and MAPE do not tell us the full story regarding the forecast variation over the year. As a matter of fact, identifying maximum error can be helpful in the heat load production planning and also can ensure the evaluation risk to cover the heating load day by day.

Figure 6 represents examples of histograms for the hourly errors for each month. The 10% and 90% quantiles are specified. In this context the width of the error's distribution varies from month to month.

In the shoulder season the forecast errors are more or less confined particularly in October of weekend model. While in the winter term the distribution errors widen specifically in January month of the weekend model. In addition, the distribution of errors is not completely symmetric around 0. The forecast errors are biased differently in each month. In our case we didn't observe any clear bias in the models.

In table 4. below a summary of the error distribution is represented. The 99% and 1% quantiles indicate the maximum errors. In winter term the best month of workdays model is December with 98% falling between -4.07 kW and 4.69 kW, while January is the worst month where 99% of errors are below 9.70 kW and 1% of data are less or equal -5.78 kW.

The best month of weekend model is February with 98% of the errors falling between -2.60 kW and 4.42 kW. while January, also here, is the worst month where 99% of errors are below 13.5 kWh and 1% are less or equal -6.13 kW.

In shoulder season the best month of workday model is October with 98% of the errors falling between -4.6 kW and 5.8 kW. In contrast April is considered as the worst month where 99% of errors are below 6.6 kW and 1% are less than -5.8 kW.

As the workday model, October represents the best month for the weekend model with 98% of errors falling between -3.4 kW and 4.5 kW and the worst model is identified in March where 99% of errors are less or equal 8.7 kW and 1% are less than -3.3 kW.

Additionally, the forecast errors are biased differently in each month for each model. In our case and according to the mean error metric (ME), we did not observe any clear bias in the four models.

## 4 Conclusion

We have presented a data driven model for the characterization and prediction of heat demand in buildings connected to district heating (DH) networks. Our new model is applied in a short-term heat load forecast with high temporal resolution (hourly and daily timestamp) for different buildings relying on weather and calendar information.

Statistically, the mean absolute percentage error (MAPE) is ranging between 4 to 20 % in winter term, which is considered as an indication that the forecast is acceptably accurate. Whereas, in shoulder term the MAPE is ranging between 20 to 80% which is considered as an indication that the forecast is low. The models in winter and shoulder term show a good performance in predicting the heat demand in early morning of the days. Having such estimation available for different buildings in district heating networks help to optimize the resource for heat generation and give rise to both energy and economic savings.

In operational forecast systems, a model with simple and reliable data input may work to make a more robust system. Having several features are not always an advantage to improve the accuracy. Nevertheless, future works should inspect different types of models to simplify the feature selection procedure and handle multiple inputs variables with noisy dependencies. Hence, make the forecasting process more applicable to a wide range of district heating systems around the world.

# Appendix 1: Summary of the hourly forecast error for each month of the four models using significant calendar data.

|  |  | RMSE | ME | Error Quantiles (kWh) | | | |
|---|---|---|---|---|---|---|---|
|  |  | (kWh) | (kWh) | 10% | 90% | 1% | 99% |
| **Winter/ Workdays model** | December | 1.865 | -0.064 | -2.238 | 2.283 | -4.079 | 4.694 |
|  | January | 3.000 | 0.055 | -2.969 | 4.092 | -5.789 | 9.702 |
|  | February | 2.339 | 0.009 | -2.321 | 2.966 | -5.240 | 6.595 |
| **Winter/ Weekends model** | December | 2.188 | -0.521 | -3.008 | 2.491 | -4.857 | 4.430 |
|  | January | 3.665 | 0.228 | -3.992 | 4.220 | -6.134 | 13.509 |
|  | February | 1.959 | 0.434 | -1.771 | 3.550 | -2.601 | 4.424 |
| **Shoulder/ Workdays model** | March | 2.142 | 0.384 | -1.893 | 2.935 | -4.979 | 6.408 |
|  | April | 2.373 | 0.005 | -2.411 | 3.478 | -5.887 | 6.619 |
|  | May | 2.359 | -0.284 | -2.799 | 2.936 | -5.201 | 5.909 |
|  | September | 2.322 | -0.226 | -2.963 | 2.486 | -4.524 | 6.371 |
|  | October | 2.029 | -0.048 | -2.175 | 2.610 | -4.605 | 5.830 |
|  | November | 2.099 | 0.225 | -1.954 | 3.319 | -3.564 | 6.237 |
| **Shoulder/ Weekends model** | March | 2.449 | 0.594 | -1.694 | 3.749 | -3.371 | 8.713 |
|  | April | 2.127 | 0.028 | -2.586 | 2.895 | -4.324 | 5.618 |
|  | May | 2.542 | -0.333 | -3.167 | 2.718 | -4.719 | 6.691 |
|  | September | 1.766 | -0.086 | -1.823 | 2.229 | -3.627 | 4.668 |
|  | October | 1.589 | -0.104 | -1.953 | 1.553 | -3.469 | 4.537 |
|  | November | 2.311 | -0.204 | -2.864 | 2.595 | -4.944 | 6.557 |